\documentclass{elsart}
\usepackage[T1]{fontenc}
\usepackage{amsmath,amssymb,mathrsfs,epsfig}
\newcommand{\mean}[1]{\langle #1\rangle}
\def\dd{\mathrm{d}}
\begin{document}
\begin{frontmatter}
\journal{Physica A}
\title{Distance-dependent connectivity:
Yet Another Approach to the Small World Phenomenon}
\author{Mat\' u\v s Medo}
\address{Department of Theoretical Physics and Physics
Education,\\
Mlynsk\' a dolina, 842 48 Bratislava, Slovak Republic}
\ead{medo@fmph.uniba.sk}
\begin{abstract}
We investigate a relationship network of humans located
in a metric space where relationships are drawn according
to a distance-dependent probability density. The obtained
spatial graph allows us to calculate the average separation
of people in a very simple manner. The acquired results agree
with the well-known presence of the small-world phenomenon
in human relationships. They indicate that this feature can
be understood merely as a consequence of the probability
composition. We also examine how this phenomenon evolves
with the development of human society.
\end{abstract}
\end{frontmatter}

\section{Introduction}
In the 1960's, the american social psychologist Stanley Milgram
examined how people know each other and introduced a quantity
named the degree of separation $D$. It is the number of people
needed to bind two chosen persons via a chain of acquaintances.
E. g., if persons $A$ and $B$ do not know each other, but they
have a common friend $C$, their degree of separation is
$D(A,B)=1$. Milgram measured the mean degree of separation
between people in USA and found a surprisingly small value,
$\mean{D}=6$. This gave another name to this phenomenon -- "six
degrees of separation".

Nowadays, if a network has small average distance between its
vertices together with a large value of the average clustering
coefficient, we say that small world phenomenon is present
and such a network is called small world network (SWN). We often
encounter SWP in random networks.

Clearly "To be an acquaintance" is a somewhat vague statement.
There are various possible definitions -- e. g. shaking the
other's hands, talking to each other for at least one hour, etc.
Fortunately, results do not depend significantly on the specific
choice, SWN was observed in all of those cases. However, the
number six in the name of the phenomenon can not be taken
literally. Actually it is just an expression for the number,
which is very small compared to the size of the investigated
population, which is taken to be $6\,400$ millions (the
approximate number of people on the Earth) in this article.

Nowadays, SWP is a well known feature of various natural and
artificial random graphs \cite{Watts}. Article citations,
World Wide Web, neural networks and other examples exhibit this
feature \cite{Dorogo-Mendes,Albert-Bar}.

There are many ways to construct a SWN. Some models are rather
mathematical and do not examine the mechanism of the origin of a
network. They impose some heuristic rules (e. g.
\cite{Erd-Ren1,Erd-Ren2,Watts-Strog}) instead. Other models look
for the reasons for the introduced rules. This is much more
satisfactory from the physicist's point of view. The first such
model is known as "preferential linking" \cite{Bar-Albert}. It
is quite reasonable for cases like the growth of the WWW, where
sites with many links to them are well known and in the future
will presumably attract more links than poorly linked pages.

In this work we focus on the random network of human
relationships. It evolves in a very complicated manner, therefore
it is very hard to impose some well accepted rules for its
growth. Hence, we do not look for the time evolution of human
acquaintances. Instead, we inquire a static case with the random
network already developed.

If the acquaintance between $A$ and $B$ is present, we link them
with an edge. We obtain the random graph of human relationships
in this way. We can introduce a metric to this network by
assigning a fixed position in the plane to every person. In
order to obtain analytical results, we assume a constant
population density. In particular, we suppose that
people--vertices are distributed regularly and form a square
lattice in the plane. With proper rescaling, the edges of unit
squares in this lattice have length $1$. Further we assume that
the probability that two people know each other, depends on their
distance by means of some distribution function. This model
should keep some basic features of the real random network of
human relationships.

\section{The Mathematical Model}
Let's have an infinite square lattice where squares have sides
equal to $1$ and there is one person in every vertex. We label
the probability that two people with distance $d$ know each other
$Q(d)$. We assume homogeneity of the population, therefore this
probability function is the same for every pair.

Summation of $Q(d)$ through all vertices leads to the average
number of acquaintances for any person which we denote
$N_{\mathrm A}$. Next, we assume that the function $Q(d)$ is
changing slowly on the scale of $1$. Therefore, we can change
summation to integration and obtain
\begin{equation}
\label{normalization}
N_{\mathrm A}=\int_{-\infty}^{\infty}\dd x
\int_{-\infty}^{\infty}\dd y\,Q\big(\sqrt{x^2+y^2}\big).
\end{equation}
Our aim is to quantify the average degree of separation
$\mean{D}$ for couples with the same geometrical distance equal
to $b$. To achieve this, we choose two such people and label
them $A$ and $B$ with positions $\vec{r}_A=[0,0]$,
$\vec{r}_B=[b,0]$ (this particular choice will not affect our
results substantially).

\section{An Analytical Solution}
Every person in the lattice can be located by its coordinates
$[x,y]$. We will denote the distance between $X$ and $Y$ as
$d_{XY}$. Let us introduce a symbol $\sim$ for the relation of
acquaintances. This is a binary relation which is symmetric but
not transitive. The probability that $X$ knows $Y$ is then
$P(X\sim Y)=Q(d_{XY})\equiv Q_{XY}$.

We name $P(D)$ the probability that the degree of separation for
$A$ and $B$ with distance $b$ is equal to the number $D$.If we
want to find out the average degree of separation in our present
network, $\mean{D}$, we need to know the probabilities $P(D)$
for all different values of $D$. At the moment, only $P(0)$ is
known, since apparently $P(0)=Q(d_{AB})=Q(b)$.
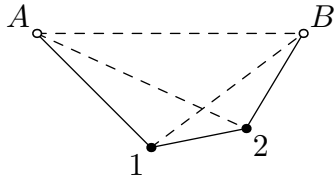
\[
\includegraphics[scale=1.2]{sw_figs.1}
\]
Let's examine the degree of separation $D=2$. This means that
there are two other persons on the path between $A$ and $B$.
We denote their coordinates as $\vec{r}_1=(x_1,y_1)$ and
$\vec{r}_2=(x_2,y_2)$. For the presence of such a track, edges
$A1$, $12$ and $2B$ are needed together with edges $A2$, $1B$ and
$AB$ missing (see picture above). Since their presence is
independent, we have
\begin{eqnarray}
\label{first}
P(2)&=&\sum_{1,2}Q_{A1}Q_{12}Q_{2B}\big(1-Q_{A2}\big)
\big(1-Q_{1B}\big)\big(1-Q_{AB}\big)\approx\notag\\
&\approx&\iint\limits_{1,2}Q_{A1}Q_{12}Q_{2B}\big(1-Q_{A2}\big)
\big(1-Q_{1B}\big)\big(1-Q_{AB}\big)\,\dd\vec{r}_1\dd\vec{r}_2.
\end{eqnarray}
where the summation runs through various placements of persons
$1$ and $2$. The change of summation to integration is possible
due to the fact that $Q(d)$ is changing slowly on the scale of
$1$.
 
Here we utilized the fact that in probabilities addition rule
$P(A\cup B)=P(A)+P(B)-P(A\cap B)$ we can neglect the last term
since probabilities $P(A)$, $P(B)$ are small and $P(A\cap B)$ is
of the higher order of smallness. Unfortunately due to this
approximation we apparently reach "probabilities" $P(D)$ higher
than $1$ for high enough value of $D$. Though probabilities
$P(D)$ small with respect to $1$ can be considered accurate.
This implies that obtained results cannot be used to evaluate
the exact value of the average degree of separation for nodes
$A$ and $B$ because in such a calculation we would need value of
$P(D)$ for every $D$. Still from the growth of $P(D)$ we can
easily see for which $D^*$ it reaches relevant values, e. g.
$P(D^*)=1/3$. This $D^*$ then characterizes the mean degree of
separation of $A$ and $B$.

We can compute the first approximation to (\ref{first}), getting
\begin{equation}
\label{1approx}
P(2)^{(0)}=\iint\limits_{1,2}Q_{A1}Q_{12}Q_{2B}
\,\dd\vec{r}_1\,\dd\vec{r}_2.
\end{equation}
As $Q_{A1}=Q(x_1-0,y_1-0)$, $Q_{12}=Q(x_2-x_1,y_2-y_1)$, and
$Q_{2B}=Q(b-x_2,0-y_2)$ we notice that (\ref{first}) is a double
convolution of the function $Q(d)$ enumerated at point $(b,0)$.
Thus we can write
\[
P(2)^{(0)}=\big[Q\ast Q\ast Q\big](b,0)\implies
P(D)^{(0)}=Q^{[D]}(b,0).
\]
For the Fourier transformation of the convolution, the following
equation holds:
\[
\mathscr{F}\big\{Q^{[D]}\big\}=\big(\mathscr{F}\{Q\}\big)^D.
\]
Using this formula we can write $P(D)^{(0)}$ in the form
\begin{equation}
\label{outcome}
P(D)^{(0)}=\mathscr{F}^{-1}
\Big\{\big(\mathscr{F}[Q]\big)^D\Big\}(b,0).
\end{equation}

The mean clustering coefficient $\mean{C}$ is the probability
that two acquaintances of $A$ know each other. It can be
evaluated in a way very similar to the calculation of
$P(D)$, the corresponding graph is on the picture below.
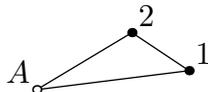
\[
\includegraphics[scale=1.2]{sw_figs.2}
\]
In order to write down an expression for $\mean{C}$ it is
straightforward to rewrite (\ref{first}). We obtain the number of
connected triples $A12$ with node $A$ fixed by this integration.
We just have to avoid double counting of every track
(interchanging positions of $1$ and $2$) -- this brings an
additional factor of $1/2$. The average number of acquaintances
for every vertex is $N_{\mathrm A}$, therefore the average number
of possible triples is
$N_{\mathrm A}(N_{\mathrm A}-1)/2\approx N_{\mathrm A}^2/2$.
The mean clustering coefficient is the ratio of the average
number of triples to the average number of possible triples.
That is,
\begin{eqnarray}
\label{cc}
\mean{C}&=&\frac1{N_{\mathrm A}^2}
\iint\limits_{1,2}Q_{A1}Q_{12}Q_{2A}
\,\dd\vec{r}_1\,\dd\vec{r}_2=
\frac1{N_{\mathrm A}^2}\big[Q\ast Q\ast Q\big](0,0)=\notag\\
&=&\frac1{N_{\mathrm A}^2}\mathscr{F}^{-1}
\Big\{\big(\mathscr{F}[Q]\big)^3\Big\}(0,0).
\end{eqnarray}

Equations (\ref{outcome}) and (\ref{cc}) are solutions of the
problem. Unfortunately, the relevant functions $Q(d)$ (see
next section) do not have an analytical form of their forward
and inverse Fourier transformation. Therefore we have to
calculate the values of $\mean{C}$ and $P(D)$ numerically.
Equation (\ref{outcome}) requires a very high calculation
precision. This makes the evaluation of $P(D)$ very slow and
even with some clever treatment (see Appendix A) it is in
practise impossible for high values of $b$. This is just our
case, because we are interested in $b=50\,000$. Thus some other
(approximate) approach is needed. First we have to find more
about the nature of function $Q(d)$.

\section{An Empirical Entries}
In the present, there are approximately $6\,400$ millions people
on the Earth. It means that length of the assumed square lattice
side is $2L=80\,000$. In order to obtain a numeric results we
choose $b=50\,000$ and the average number of acquaintances
$N_{\mathrm A}=1\,000$.

To get some insight on the distribution $Q(d)$, some analysis is
needed. First it is clear that $Q(d)$ should be decreasing with
$d$. Moreover, closely living people know each other almost
certainly. That is
\begin{equation}
\label{limit}
\lim_{d\to0} Q(d)=1.
\end{equation}
Together with (\ref{normalization}) we now have two requirements
for $Q(d)$. Indeed, there are many functions satisfying them,
e. g. we can choose $Q(d)=C\exp[-r/a]$.

The last quantity we can compute is the average number of
{\em distant people\/} every person in the lattice know,
$N_{\mathrm d}$. Here {\em distant\/} means that people's
distance from the chosen fixed person (node) is greater than
$L/2$. This is a simple analogy to the number of people we know
on the other side of the Earth. So we have
\begin{equation}
N_{\mathrm d}=N_{\mathrm A}-2\pi\int\limits_0^{L/2}rQ(r)\,\dd r.
\end{equation}
If exponential distribution discussed above satistfies
(\ref{normalization}) and (\ref{limit}) it folllows that
$N_{\mathrm d}\approx10^{-13}$. This is in a clear contradiction
to the fact that there are people who have very distant friends.
Still we can improve $N_{\mathrm d}$ if we use stretched exponential
$Q(d)=\exp[-K\,d^a]$ with exponent $a$ between $0.2$ and
$0.3$.\footnote{Approximate solution presented in next section
can be used also for this distribution.}
However, if we check $Q(1)$ (probability to know our closest person)
it is well below $0.3$. Stretched exponentials therefore satisfy
condition (\ref{normalization}) just formally and we will not it
discuss it later. Moreover, mean clustering coefficient is then
very small (from $2\cdot 10^{-4}$ to $3\cdot10^{-3}$).

Now it's clear that distribution $Q(d)$ can't decrease so fast as
exponential functions, wide tails are inevitable in our model.
This leads us to power-law distributions $1/x^a$. According to
(\ref{limit}) we demand
\begin{equation}
\label{powerlaw}
Q_a(d)=\frac1{1+bd^a},
\end{equation}
where $b$ is fixed by (\ref{normalization}). Number of far
friends now ranges from $N_{\mathrm d}\approx0.01$ ($a=3.5$)
to $N_{\mathrm d}\approx 17$ ($a=2.5$). This range of exponents
gives us reasonable range for values of $N_{\mathrm d}$.

In this article we also show results for the normal distribution
$Q_{\mathrm n}(d)=\exp[-ad^2]$ ($N_{\mathrm d}\approx0$)
and the uniform distribution within fixed radius
$Q_{\mathrm u}(d)=\vartheta(R_{\mathrm A}-d)$ ($N_{\mathrm d}=0$).

With regard to the fact that all used distributions $Q(d)$
approach to zero for large values of $d$ it is almost certain
that the shortest chain of acquaintances between chosen $A$ and
$B$ do not run out of the examined lattice with side $80\,000$.
Therefore it doesn't matter if we have integration (summation)
bound in infinity or $\pm L=\pm 40\,000$. This allows us to
use all results derived for infinite lattice in the real case
of finite lattice.

\section{An Approximate Solution for Power-law Distributions}
To demonstrate the calculation we take $P(2)$ as an example
again. In the previous section we found out that power-law
distributions are especially important in our model. Their joint
probability $Q(r_1)Q(b-r_1)$ has sharp maximum for $r_1=0$
and low minimum for $r_1=b/2$. Their ratio is
\[
\frac{Q(b/2)^2}{Q(b)Q(0)}\approx\Big(\frac{4}{b}\Big)^a
\]
where $a$ is the exponent in (\ref{powerlaw}). This implies that
in (\ref{first}) we can constrain summation to
$r_{A1},r_{A2}\ll b$ or $r_{B2},r_{B1}\ll b$ or
$r_{A1},r_{B2}\ll b$ (see picture below).
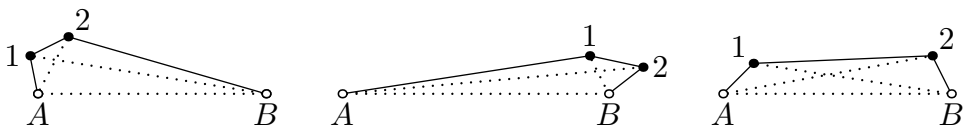
\[
\includegraphics[scale=1.2]{sw_figs.3}
\]
Here we obtained three different diagrams. Let's examine first
one in detail.

Since edges $AB$ and $B1$ are long we can write
\[
P(2)\approx\iint\limits_{1{,}2}
Q_{A1}Q_{12}Q_{2B}\big(1-Q_{A2}\big)
\,\dd\vec{r}_1\,\dd\vec{r}_2.
\]
It is easy to show that for power-law distributions
$Q(b-r_1)\approx Q(b)\equiv Q_{AB}$ when $r_1\ll b$.
Thus we have (for corresponding diagram see picture below)
\begin{align*}
P(2)&\approx\iint\limits_{1{,}2}
Q_{A1}Q_{12}Q_{AB}\big(1-Q_{A2}\big)
\,\dd\vec{r}_1\,\dd\vec{r}_2=\\
&=Q(b)\iint\limits_{1{,}2}
Q_{A1}Q_{12}\,\dd\vec{r}_1\,\dd\vec{r}_2-
Q(b)\iint\limits_{1{,}2}
Q_{A1}Q_{12}Q_{A2}\,\dd\vec{r}_1\,\dd\vec{r}_2.
\end{align*}
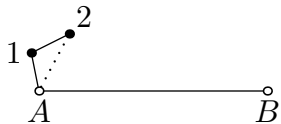
\[
\includegraphics[scale=1.2]{sw_figs.4}
\]
Both integrals are easy to compute. Second one brings average
clustering coefficient $\mean{C}$ into account. The result is
\[
P(2)\approx Q(b)N_{\mathrm A}^2\big(1-\mean{C}\big).
\]
Remaining two diagrams for $P(2)$ can be evaluated in the same
way.

In the computation of $P(D)$ for higher values of $D$ we
encounter products of kind $(1-Q_{13})(1-Q_{24})\ldots$ even
after neglecting probabilities $Q_{ij}$ for long edges $ij$.
Here we can make first order approximation
\[
(1-Q_{13})(1-Q_{24})\approx 1-Q_{13}-Q_{24}
\]
which is valid almost everywhere except small spatial region
that do not contributes substantially (see section Results and
discussion). Moreover, second approximation considering
terms $Q_{13}Q_{24}$ would increase evaluated probabilities.
Therefore first approximation results will be some lower bound
estimates of $P(D)$.

Higher values of $D$ introduce long closed loops of kind
$A12\ldots nA$ ($n\leq D$). Corresponding integrals can be
carried out in the same way like it was presented in the
derivation of (\ref{cc}). Finally we obtain
\begin{equation}
\label{cn}
C_n\equiv\frac{1}{N_{\mathrm A}^n}\iint\limits_{1{,}2}
Q_{A1}Q_{12}\cdots Q_{nA}\,\dd\vec{r}^n=
\frac{1}{N_{\mathrm A}^n}\mathscr{F}^{-1}
\Big\{\big(\mathscr{F}[Q]\big)^n\Big\}(0,0).
\end{equation}
This helps us to find values of $C_n$ for any $n$. Clearly
$C_2=\mean{C}$. With the use of such a closed loop integrals
we can write
\begin{equation}
\label{final}
\begin{aligned}
P(0)&=Q(b),\\
P(1)&=Q(b)N_{\mathrm A}2,\\
P(2)&=Q(b)N_{\mathrm A}^2\big(3-2C_2\big),\\
P(3)&=Q(b)N_{\mathrm A}^3\big(4-6C_2-2C_3\big),\\
P(4)&=Q(b)N_{\mathrm A}^4\big(5-12C_2-6C_3-2C_4\big),\\
P(5)&=Q(b)N_{\mathrm A}^5
\big(6-20C_2-12C_3-6C_4-2C_5\big),\,\ldots
\end{aligned}
\end{equation}

\section{Results and Discussion}
In this section we summarize results for various distributions
$Q(d)$ ranging from the uniform $Q_{\mathrm u}$ and normal
$Q_{\mathrm n}$ to power-law distributions $Q_{3.5}$--$Q_{2.5}$
(see (\ref{powerlaw})) and flat distribution $Q_{\mathrm ER}$.
This list is sorted according to the quantity of long shortcuts
in such networks of relationships.

\subsection*{Flat Distribution}
If we have flat distribution $Q_{\mathrm ER}$, every pair of
vertices is connected with the same probability $p$. It is shown
in \cite{Erd-Ren1} that in the network consisting of $N$ vertices
holds
\[
\mean{l}\approx\frac{\ln N}{\ln pN}.
\]
Here $pN$ is the average number of acquaintances for a person in
the network, $pN=N_{\mathrm A}$. We have $N_{\mathrm A}=1\,000$
and $N=6\,400$ millions thus
$D^*_{\mathrm ER}=\mean{l}-1\approx 2.3$ and $\mean{C}\approx0$.

\subsection*{Uniform Distribution Within Fixed Radius}
We can discuss such a case where every person knows just
$N_{\mathrm A}$ closest neighbors. This leads us to the distribution
$Q_{\mathrm u}(d)=\vartheta(R_{\mathrm A}-d)$ where distance
$R_{\mathrm A}$ is fixed by (\ref{normalization}). It gives us
$R_{\mathrm A}=\sqrt{N_{\mathrm A}/\pi}$ and therefore
\[
D^*_{\mathrm u}\approx b\,\sqrt{\frac{\pi}{N_{\mathrm A}}}.
\]
It's worth to note that we don't have any randomness in this
model thus
$D^*_{\mathrm u}=\mean{D_{\mathrm u}}$.\footnote{Randomness
can be introduced by random placement of vertices. Hence we
obtain
so called random geometric graphs discussed in \cite{Penrose}. This
approach is complementary to presented one where vertex placement is
fixed but their connecting is due to some probability distribution.}

\subsection*{Normal Distribution}
The only distribution which allows us to evaluate (\ref{outcome})
analytically is normal distribution $Q_{\mathrm n}$. The
result is
\[
P(D)=\frac{N_{\mathrm A}^D}{D+1}\exp
\bigg[-\frac{\pi b^2}{N_{\mathrm A}(D+1)}\bigg].
\]
It was argued before that a solution of the equation
$P(D^*_{\mathrm n})=1/3$ characterizes value of the mean degree
of separation $\mean{D}_{\mathrm n}$. For $N_{\mathrm A}=1\,000$
and $b=50\,000$ we can use some approximations which lead us to
\[
D^*_{\mathrm n}\approx n_{\mathrm n}\approx
b\,\sqrt{\frac{\pi}{N_{\mathrm A}\ln N_{\mathrm A}}}=
\frac{D^*_{\mathrm u}}{\sqrt{\ln N_{\mathrm A}}}.
\]
The actual value of $D^*_{\mathrm n}$ is about one third of
$D^*_{\mathrm u}$ (this is due to the existence of some
longer connections in the network, although it is extremely
suppressed by the exponential decay). We can note that both
$D^*_{\mathrm u}$ and $D^*_{\mathrm n}$ scale with $b^1$. This
clearly differs from $\ln b$ scaling of the Erd\" os-R\' enyi
model. The clustering coefficient $\mean{C}$ can be evaluated
easily both for normal and uniform distribution. We obtain high
values of $\mean{C}$ (see graph below) in both cases. This
agrees with our expectations.

\subsection*{Power-law Distributions}
\begin{figure}[b]
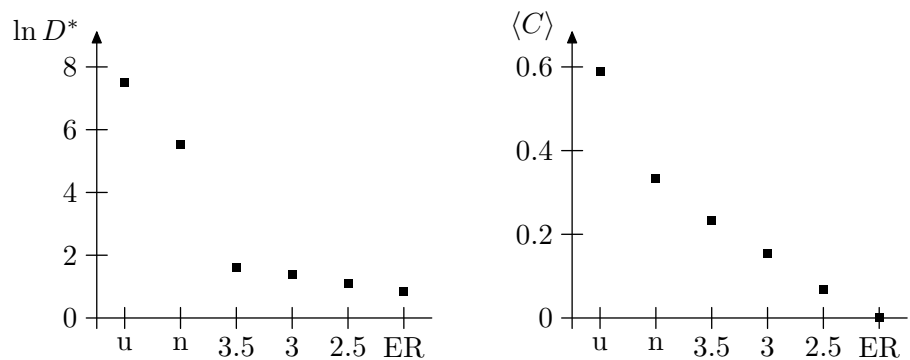

\begin{center}
\includegraphics[scale=1.1]{sw_figs.5}
\qquad
\includegraphics[scale=1.1]{sw_figs.6}
\end{center}
\caption{Graphs of mean degree of separation and clustering
coefficients for various distribution functions.}
\label{fig:graf}
\end{figure}

Numerical computation of coefficients $C_2,\ldots,C_5$
with (\ref{cn}) is rather fast -- their values are shown
in the table below.
\begin{center}
\begin{tabular}{|c|c|c|c|}
\hline
& $a=2.5$ & $a=3.0$ & $a=3.5$\\
\hline
$C_2$ & 0.068 & 0.154 & 0.233\\
$C_3$ & 0.030 & 0.090 & 0.153\\
$C_4$ & 0.016 & 0.059 & 0.109\\
$C_5$ & 0.009 & 0.042 & 0.084\\
\hline
\end{tabular}
\end{center}
Substituting these values into (\ref{final}) leads us to
values of mean degree of separation which are marked in the
Fig. \ref{fig:graf}. We see that power-law distributions
$Q(d)$ results in high values of mean clustering coefficients
$\mean{C}=C_2$ together with small values of $D^*$ (from $6$
to $4$). Thus small world phenomenon is clearly present in
these networks.

One can also ask for some comparison with the well-known
Barabasi-Albert model. Mean vertex degree is then
$\mean{k}=2m$ and mean clustering coefficient is
$\mean{C}=(m-1)\ln^2N/8N$ (here $m$ is degree of just
added vertices, see \cite{FFH}). With respect to our choice
$N_{\mathrm A}=1\,000$, $N=6.4\cdot 10^9$ it follows that
$m=500$ and $\mean{C}\approx 10^{-11}$. This is certinaly
nonrealistic value, our model gives better estimation
of $\mean{C}$.

For presented values of coefficients $C_i$ expressions in
parentheses in (\ref{final}) do not fall close to zero for
quite wide range of values of $D$. Therefore we can (very
approximately) write
\[
P(D)\approx Q(b)DN_{\mathrm A}^D.
\]
Solution of the equation $P(D^*)=1/3$ is approximately
$D^*\approx a\ln b/\ln N_{\mathrm A}$. With $b$ this scales as
$\ln b$. This is very different from $b^1$ scaling of $\mean{D}$
for the uniform a normal distribution. Such a scaling is similar
to the scaling in the Erd\" os-R\' enyi model, though values of
clustering coefficient are kept high as we demanded in the
introduction.

Probability $P(2)$ can be evaluated also by straightforward
summation in accordance with (\ref{first}) although it takes
huge amount of computer time. Obtained values agree very well
with results presented above for all examined exponents but
$2.5$ -- this case requires more computer time than it was
given. Computation of $P(3)$ in the same way exceeds our
computer possibilities for every exponent but we do not
regard it necessary.

\subsection*{Time Evolution and Some Limitations}
Human relationships in modern world are much more widespread
than it was in the past. One can think of slowly changing
exponent of the power-law distribution function $Q(d)$ from large
values to smaller (perhaps resulting to almost flat distribution
in the future -- internet helps to bridge the distances). According
to the Fig. \ref{fig:graf} we see that this would affect exact
value of clustering coefficient. However it would remain high
enough for wide range of exponents. Similarly changes of mean
degree of separation are not important at all -- it remains very
small compared to the size of human population.

Finally it has to be noted that in the described model we do not
consider presence of some organized hierarchic structures in
human society. E. g. chief of the firm knows his employees, but
he also knows another chiefs who know their employees, etc.
Amount of people involved in the hierarchical tree grows
exponentially with the number of its levels. Such an arrangement
therefore introduces additional way how to know each other with
small resulting degree of separation. In presented calculation we
didn't include this effect. Yet there is one important insight.
If we proved the degree of separation being small without
considering of the hierarchies, their presence would even
decrease it.

\section{Conclusion}
We have examined the mean degree of separation and the
clustering coefficient for a random network of human
relationships in this article. We were able to compute these
quantities in our model. For a power-law decay of probability
$Q(d)$, we obtained a small mean degree of separation compared
to the size of the network, along with a large value of the mean
clustering coefficient. Both of these features are typical for
small world networks. Thus we have shown that the small world
phenomenon can be understood as a simple consequence of
additivity of probabilities.

We saw that the style of calculation depends on the used
distribution $Q(d)$. The computation was finished analytically
for some special cases. In other cases, thanks to some
approximations, we utilized the advantage of (\ref{cc}) where $b$
do not enter the inverse Fourier transformation, making it easy
to evaluate numerically.

It's worth to note that the model solved herein is similar to the
Watts and Strogatz model \cite{Watts-Strog} where long shortcuts
were introduced by a random rewiring procedure. In our model long
shortcuts are present thanks to wide tails of power-law
distributions. This model model brings two basic advantages. First,
the derivation and the resulting relations for $C$ and $D^*$ are
more simple. Moreover, our model has more realistic foundations. Nevertheless, the typical behavior of this model is the same as
in previous models. The introduction of long shortcuts to the system
decreases the average degree of separation rapidly, but also keeps
the clustering coefficient high enough for the so called small world
phenomenon to appear.

\appendix
\section{Numerics of the Fourier Transformation}
The Fourier integrals encountered in the solution of presented
problem can not be solved analytically thus numerical techniques
have to be used. In the inverse Fourier transform this is
especially awkward because we meet rapidly oscillating term
$\exp[\mathrm{i}\,bu]$. Here $b$ is the distance
between chosen persons $A$ and $B$, by assumption big number
($b=50\,000$). Therefore we have to compute Fourier
transformation of $f(d)$ very accurately. In order to make
computation less demanding on the computer time, it is convenient
to find some approximation in the computing of the inverse
Fourier transformation. We will continue with this derivation in
the onedimensional case for the sake of simplicity.

The Fourier transformation of the even function $f(x)$ is an even
real function. According to the (\ref{outcome}) we are looking
for the inverse Fourier transformation of its $n$-th power, we
will denote it $\hat{g}(u)$. It is also even real function.
Therefore its inverse Fourier transformation is real function
(sine-proportional terms vanish). Thus
\[
g(b)=\frac1{2\pi}\int\limits_{-\infty}^{\infty}\hat{g}(u)
\mathrm{e}^{\mathrm{i}bu}\,\dd u=\frac1{2\pi}
\int\limits_{-\infty}^{\infty}\hat{g}(u)\cos[bu]\,\dd u.
\]
This integral can be expressed as the sum of
contributions from all periods of the $\cos[bu]$ function,
$I_n=\mean{2\pi n/b,2\pi(n+1)/b}$ (here $n\in\mathbb{N}$)
\[
g(b)=\sum_{n=-\infty}^{\infty} S_n(b),\quad
S_n(b)=\frac1{2\pi}\int\limits_{I_n}\hat{g}(u)\cos[bu]\,\dd u.
\]
In the integrand of previous equation we can make Taylor
expansion of $\hat{g}(u)$ around $\xi_n=2\pi(n+1/2)/b$.
Thereafter terms of kind $u^m\cos[bu]$ emerge ($m\in\mathbb{N}$).
Such integrals are easy to compute -- first two terms of
resulting expansion are then
\[
S_n(b)=\frac1{b^3}
\frac{\dd^2\hat{g}}{\dd u^2}\bigg\rvert_{\xi_n}+
\frac{\pi^2-6}{6b^5}
\frac{\dd^4\hat{g}}{\dd u^4}\bigg\rvert_{\xi_n}.
\]
Finally we have
\begin{equation}
\label{priblizenie}
g(b)=\frac1{b^3}\sum_{n=-\infty}^{\infty}
\frac{\dd^2\hat{g}}{\dd u^2}\bigg\rvert_{\xi_n}+
\frac{\pi^2-6}{6b^5}\sum_{n=-\infty}^{\infty}
\frac{\dd^4\hat{g}}{\dd u^4}\bigg\rvert_{\xi_n}.
\end{equation}
This helps us to speed up inverse Fourier transformation -- we
do not have to know so many values of $\hat{g}(u)$. For every
range $I_n$ evaluation of $\hat{g}(u)$ in three points (for
numerical calculation of second derivative in the leading term of
(\ref{priblizenie})) is enough. We just have to keep in mind that
these points have to be close enough (with respect to $2\pi/b$),
otherwise we can obtain evidently incorrect results (e. g.
$g(b)=0$ when border points have distance $2\pi/b$).

\begin{ack}
The author would like to thank to staff of his department,
especially to Martin Moj\v zi\v s and Vladim\' ir \v Cern\' y
for valuable conversations and to Mari\' an Klein for computer
time. Acknowledgement belongs also to J\' an Bo\v da for
introduction to the field, Mi\v ska Sonlajtnerov\' a for her
enthusiastic encouragement and my parents for their support.
\end{ack}

\end{document}